\documentclass[%
 aip,
 amsmath,amssymb,
 reprint,%
]{revtex4}

\usepackage{graphicx}% Include figure files
\usepackage{dcolumn}% Align table columns on decimal point
\usepackage{bm}% bold math
\usepackage[utf8]{inputenc}
\usepackage[T1]{fontenc}
\usepackage{mathptmx}
\usepackage{etoolbox}

\makeatletter
\def\@email#1#2{%
 \endgroup
 \patchcmd{\titleblock@produce}
  {\frontmatter@RRAPformat}
  {\frontmatter@RRAPformat{\produce@RRAP{*#1\href{mailto:#2}{#2}}}\frontmatter@RRAPformat}
  {}{}
}%
\makeatother
\begin{document}

\preprint{AIP/123-QED}

\title[Sample title]{Comment on "On the relations between large-scale models
of superfluid helium-4" [Phys. Fluids 33, 127124 (2021]"}
%\\with Forced Linebreak}
% Force line breaks with \\
\author{Sergey K. Nemirovskii}
 \altaffiliation {Institute of Thermophysics, Lavrentyev ave, 1, 630090, Novosibirsk, Russia}
 %%Lines break automatically or can be %forced with \\
%\author{B. Author}%
 \email{nemir@itp.nsc.ru}

\date{\today}% It is always \today, today,

\begin{abstract}
We comment on the paper by M. Sýkora, M. Pavelka, M. La Mantia, D. Jou, and M. Grmela
 "On the relations between large-scale models of superfluid helium-4,"
Physics of Fluids, 33(12):127124(2021),where the authors  have developed a formalism for describing a coarse-grained flow
of superfluid helium. This formalism is greatly based on the 
Hall-Vinen-Bekarevich-Khalatnikov (HVBK) model. We strongly disagree with
the use of the HVBK equation approach for the case of the three-dimensional
quantum turbulence and expose our objections in this comment. We discuss the
HVBK method and also criticize the so-called vortex bundles model, which
serves as a basis for using the HVBK method in a three-dimensional quantum
turbulence.
\end{abstract}

\maketitle

%\begin{quotation}
%The ``lead paragraph'' is encapsulated with the \LaTeX\
%\verb+quotation+ environment and is formatted as a single paragraph before the first section heading.
%(The \verb+quotation+ environment reverts to its usual meaning after the first sectioning command.)
%Note that numbered references are allowed in the lead paragraph.
%
%The lead paragraph will only be found in an article being prepared for the journal \textit{Chaos}.
%\end{quotation}

%\section{\label{sec:level1}First-level heading:\protect\\ The line
%break was forced \lowercase{via} \textbackslash\textbackslash}

In the paper \cite{Jou2022}, the authors the authors have developed a formalism for describing 
a coarse-grained flow
of superfluid helium. This formalism includes a description of mutual
friction between the normal and superfluid components due to quantum
vortices. Their approach is based on the so called
Hall-Vinen-Bekarevich-Khalatnikov (HVBK) model. We disagree with
the use of this model to the arbitrary three-dimensional
flow, particularly, to the quantum turbulence, and expose our objections in
this comment. We critically discuss the HVBK method and claim that vortices should be
taken into account within  a more general formalism.

For the sake of transparency we discuss further in terms
of the standard two-fluid model, dealing with the normal and superfluid
velocities and density and entropy fields, as the more generally accepted
approach. This doesn't affect the question of validity (unvalidity) of  the HVBK
approach to an arbitrary flow of superfluids.

The flow of superflluids in the presence of
quantized vortices (e.g. quantum turbulence) is a very deep physical phenomenon, that has
several levels for its understanding. Probably the lowest of them is the so
called coarse-grained description, when the entire (infinite) set of the
vortex tangle characteristics is represented by a single variable - the
vortex line density (VLD) $\mathcal{L}(r,t)$ - the total length of lines per
unit volume. In practice, some additional reasonings about the structure of the
vortex tangle i.e. isotropy (or anisotropy), polarization, etc. is required,
but usually these questions are more or less resolved. Despite such
a rough approximation, this approach turns out to be very valuable for
macroscopic problems, primarily for hydrodynamical and  engineering tasks.

Hydrodynamic equations for the coarse-grained flow of superfluid helium read (
for details and notations see \cite{Nemirovskii2013} )
\begin{equation}
\rho _{n}{\frac{\partial \mathbf{v}_{n}}{\partial t}}+\rho _{n}(\mathbf{v}%
_{n}\cdot \nabla )\mathbf{v}_{n}=-{\frac{\rho _{n}}{\rho }}\nabla p_{n}-\rho
_{s}S\nabla T+\mathbf{F}_{mf}+\eta \nabla ^{2}\mathbf{v}_{n},
\label{eq norm}
\end{equation}%
\begin{equation*}
\rho _{s}{\frac{\partial \mathbf{v}_{s}}{\partial t}}+\rho _{s}(\mathbf{v}%
_{s}\cdot \nabla )\mathbf{v}_{s}=-{\frac{\rho _{s}}{\rho }}\nabla p_{s}+\rho
_{s}S\nabla T-\mathbf{F}_{mf}..
\end{equation*}%

The action of quantum vortices is reduced to the friction force $\mathbf{F}_{mf}$ between
the superfluid and normal components. This friction  leads to
a striking difference between the usual two fluid hydrodynamics and motion in
presence of quantum vortices. Therefore knowing the value of $\mathbf{F}%
_{mf}$ is crucial to developing a correct theory. The generally
accepted way is to expression the mutual friction $%
\mathbf{F}_{MF}$ is the following:%
\begin{equation}
\mathbf{F}_{MF}=B(T,p)\mathcal{L}(\mathbf{r},t)(\mathbf{v}_{n}-\mathbf{v}%
_{s}).  \label{Fmf macro}
\end{equation}

The coefficient $B(T,p)$ depends on the friction coefficient and
structure of the vortex tangle. in a steady counterflow $B(T,p)$ is
proportional to the famous Gorter - Mellink constant $A(T,p)$ \cite{Nemirovskii2013}.
Equations (\ref{eq norm}) - (\ref{Fmf macro}) are  the consensus point
among physicists. The controversy begins with the question of how to treat the quantity $\mathcal{%
L}(\mathbf{r},t)$, included in (\ref{Fmf macro})- the so called closure
procedure. In papers  \cite{Nemirovskii2013},\cite%
{Nemirovskii2020} an analysis of several approaches for
the closure procedure is presented.

The authors of the article \cite{Jou2022} used a closure procedure based on the HVBK model,
initially developed
for the hydrodynamics of rotating helium (see, e.g. \cite{Khalatnikov1965}).
The essense of this approach
is to exclude the VLD $\mathcal{L}(r,t)$ using the ansatz
\begin{equation}
\left\vert \nabla \times \mathbf{v}_{s}\right\vert =\kappa \mathcal{L}.
\label{curl v}
\end{equation}%
Ansatz (\ref{curl v}) is related to the famous Feynman rule
on the number of vortex lines in a rotating He II.  My objection is to the use of
the HVBK in the case of an arbitrary three-dimensional flow of He II, which is studied
in \cite{Jou2022}. This approach was designed and
applicable only for the case of stationary rotating helium and weak
deviations from it (see e.g. \cite{Sonin2016}). In other cases it is not applicable.
Of course, in the presence of quantum vortices, the coarse-grained vorticity of superfluid component
$\left\vert \nabla \times \mathbf{v}_{s}\right\vert$ may be nonzero, however it is
not directly connected to the VLD $\mathcal{L}(r,t)$ via
any simple relation like Eq. ((\ref{curl v})). Other, more subtle characteristics of the vortex tangle must be
involved. But then the problem falls out of coarse-grained hydrodynamics.  As an
example, we point out the counterflowing (Vinen) turbulence when
both VLD $\mathcal{L}(r,t)$ and friction force $\mathbf{F}_{MF}$ exist, whereas $%
\left\vert \nabla \times \mathbf{v}_{s}\right\vert =0$.
 The HVBK procedure described above was used in article \cite{Jou2022},
(see formula (32e) and text after it). The authors directly obtained that the
mutual friction $\mathbf{F}_{mf}$ is proportional to the vorticity $
\left\vert \nabla \times \mathbf{v}_{s}\right\vert$,
which is consistent with the (\ref{curl v}).

Anticipating objections of the authors I would like to discuss
usual arguments in favor of using the HVBK method in three-dimensional
flows. In fact, there is only one argument suggesting that the vortex
tangle consists of so-called vortex bundles, which contain ALL vortex
filaments, existing in the vortex tangle (otherwise, the use of the Feynman
rule  (\ref{curl v}) is invalid) (see e.g. \cite{Baggaley2012e}).
However,  this model is extremely unrealistic, vortex bundles, even those created
artificially, are destroyed very quickly, for example, due to reconnections. Critical analysis
of the concept of vortex beams was carried out in the works\cite{Nemirovskii2013},%
\cite{Nemirovskii2020}.

A somewhat mysterious question arises: how happened that the HVBK method,
designed for rotational cases, began to be used for
three-dimensional turbulent flows? The authors of \cite{Jou2022} refer to a
number of papers (Refs. [18]-[20]). But the first two works deals with the
rotating He II. The third paper \cite{Holm2001} started with the text:
"Recent experiments establish the Hall-Vinen-Bekarevich-Khalatnikov (HVBK)
equations as a leading model for describing superfluid Helium turbulence.
See Nemirovskii and Fiszdon [1995] and Donnelly [1999] for authoritative
reviews." But R. Donnelly \cite{Donnelly1999} discussed HVBK approach specifically
for rotating helium. As for my (with W. Fiszdon) paper \cite{Nemirovskii1995}
on superfluid turbulence, firstly, there was no mention of the HVBK theory
at all, and, secondly, I generally opposed this method to study
three-dimensional quantum turbulence. Thus, the origin of the idea of
using a pure rotational HVBK approximation for a three-dimensional turbulent
flows is rather vague.
    The only advantage of the HVBK approach is to get rid of  (by incorrect
use of HVBK ansatz) the dynamics of vortex filaments, and to reduce
everything to the familiar Navier-Stokes type equations. However, in my sight,
this is not an advantage, but, on the contrary, it is great disadvantage, because
that an extremely fascinating and interesting phenomenon - quantum turbulence
is completely excluded during the HVBK procedure. This seems counterproductive,
since, after the pioneering works of Feynman, Vinen, Donnelly, Schwarz and others,
it was common to imagine the vortex tangle as a set of stochastic loops
with rich and diverse dynamics, which is completely ignored in the HVBK approach.

    For the reasons stated above, I argue that HVBK approach, used in paper
\cite{Jou2022} and the ansatz $\left\vert \nabla \times \mathbf{v}%
_{s}\right\vert =\kappa \mathcal{L}$ associated with it, cannot be
applied for description of flows with mutual friction (except rotation), so
the corresponding results are highly  questionable. I think that the
description of the course-grained flows of superflluids in the presence of
quantized vortices should be carried out on the basis of a more general formalism
in which the VLD $\mathcal{L}(\mathbf{r},t)$ is considered as an
equivalent independent variable. An example of such an approach is the
Hydrodynamics of Superfluid Turbulence (HST) reviewed
in \cite{Nemirovskii1995}, \cite{Nemirovskii2013},\cite{Nemirovskii1983}.

In conclusion, I would like to add that the HVBK method can potentially be
applied in cases other than pure rotation. However, any such extenstion requires
a separate rigorous justification. As an example we refer to studies
of the Taylor–Couette flow \cite{Henderson1995a}.

The various aspects of the author's activity on the topic discussed in this
paper were fulfilled within the framework of the State Task at the Institute
of Thermophysics of the Siberian Branch of the Russian Academy of Sciences
(AAAA-A17-117022850027-5), and Russian Science Foundation (Project No.
18-0800576).
%\bibliographystyle{apsrev4-1}

%\bibliographystyle{unsrt}
%\bibliography{E:/AASERGEY/PAPERS/1BIBTEX/QT}

%\bibliography{aipsamp}% Produces the bibliography via BibTeX.

\end{document}